\documentclass[acmsmall]{acmart}

\usepackage{tabularx}
\usepackage{hyperref}
\usepackage{CJK}
\usepackage{rotating}
\usepackage[normalem]{ulem}

\useunder{\uline}{\ul}{}
\usepackage{enumitem}

\newcommand{\placeholder}[1]{\textcolor{red}{#1}}

\AtBeginDocument{%
  }

\setcopyright{cc}
\setcctype{by}
\acmJournal{PACMHCI}
\acmYear{2026} \acmVolume{10} \acmNumber{2} \acmArticle{CSCW030}
\acmMonth{4} \acmPrice{} \acmDOI{10.1145/3788066}

\begin{document}

\title{``My Body is Not Your Porn'': Identifying Trends of Harm and Oppression through a Sociotechnical Genealogy of Digital Sexual Violence in South Korea}
\renewcommand{\shorttitle}{Identifying Trends of Harm and Oppression in Digital Sexual Violence in South Korea}

\author{Inha Cha}
\authornote{Author names are in alphabetical order. Authors contributed equally to this research.}
\email{icha9@gatech.edu}
\orcid{0009-0004-2228-2413}
\affiliation{
    \institution{Georgia Institute of Technology}
    \city{Atlanta}
    \state{Georgia}
    \country{USA}
}

\author{Yeonju Jang}
\authornotemark[1]
\email{yj376@cornell.edu}
\orcid{0000-0003-1606-874X}
\affiliation{
    \institution{Cornell University}
    \city{Ithaca}
    \state{New York}
    \country{USA}
}

\author{Haesoo Kim}
\authornotemark[1]
\email{hk778@cornell.edu}
\orcid{0000-0002-7028-4294}
\affiliation{
    \institution{Cornell University}
    \city{Ithaca}
    \state{New York}
    \country{USA}
}

\author{Joo Young Park}
\authornotemark[1]
\email{jooyoung@kth.se}
\orcid{0000-0002-0767-6973}
\affiliation{
    \institution{KTH Royal Institute of Technology}
    \city{Stockholm}
    \country{Sweden}
}

\author{Seora Park}
\authornotemark[1]
\email{seorpark@iu.edu}
\orcid{0000-0001-7281-4538}
\affiliation{
    \institution{Indiana University Bloomington}
    \city{Bloomington}
    \state{Indiana}
    \country{USA}
}

\author{EunJeong Cheon}
\authornote{Corresponding author.}
\email{echeon@syr.edu}
\orcid{0000-0002-0515-6675}
\affiliation{
    \institution{Syracuse University}
    \city{Syracuse}
    \state{New York}
    \country{USA}
}

\begin{abstract}
Ever since the introduction of internet technologies in South Korea, digital sexual violence (DSV) has been a persistent and pervasive problem. Evolving alongside digital technologies, the severity and scale of violence have grown consistently, leading to widespread public concern. In this paper, we present four eras of image-based DSV in South Korea, spanning from the early internet era of the 1990s to the deepfake scandals in the mid-2020s. Drawing from media coverage, legal documents, and academic literature, we elucidate forms and characteristics of DSV cases in each era, tracing how entrenched misogyny is reconfigured and amplified through evolving technologies, alongside shifting legislative measures. Taking a genealogical approach to read prominent cases of different eras, our analysis identifies three constitutive and interconnected dimensions of DSV: (1) the homo-social fabrication of ``obscenity'', wherein victims' imagery becomes collectively framed as obscene through participatory practices in male-dominant networks; (2) the increasing imperceptibility of violence, as technologies foreclose victims' ability to perceive harm; and (3) the commercialization of abuse through decentralized economic infrastructures. We suggest future directions for CSCW research, and further reflect on the value of the genealogical method in enabling non-linear understanding of DSV as dynamically evolving sociotechnical configurations of harm. 

\placeholder{Warning: This paper includes descriptions and mentions of sexual violence, including violence towards minors and children.}
\end{abstract}



\received{May 2025}
\received[revised]{November 2025}
\received[accepted]{December 2025}


\maketitle
\begin{CJK}{UTF8}{mj}

\section{Introduction}

\begin{quote}
\textit{``How much longer will men treat women as mere objects of sexual desire? Women's lives are not your pornography! Punish the creators, distributors, and viewers ALIKE!''}~\cite{hae-rin_2024}
\end{quote}

Enraged cries echoed throughout the streets of Seoul, the capital city of South Korea, September 2024. Tens of thousands of protesters, most of them women, lined the streets to protest against the pervasive culture of digital sexual violence (DSV)~\cite{hankyoreh2024}. These were scenes from a powerful demonstration that unfolded in front of Hyehwa Station in Seoul, near Daehak-ro, a prominent cultural and artistic district. This striking scene was reminiscent of the 2018 protests, the first of a series of public demonstrations and protests where women rallied against spycam crimes (known in South Korea as \textit{Molka}). Six years later, the banners changed, but the anger remained. The Joint Action to Condemn Misogynistic Violence, a coalition of students from six women’s universities in Seoul, organized a protest that drew 6,000 women from across the country. Chanting ``Punish creators, distributors, and viewers alike'', demonstrators filled Daehak-ro. They condemned lawmakers for failing to enact effective legislation, courts for lenient punishments, and police for their lack of investigative will~\cite{hae-rin_2024, hankyoreh2024}.

This outcry was motivated by the recently-uncovered epidemic of deepfake sexual exploitation and abuse, which had garnered widespread attention earlier that year. In May 2024, news broke of systematic networks of Telegram chat rooms used to generate and share sexual imagery of women and young girls ~\cite{mackenzie2024deepfake,joongang_2024}. News reported of perpetrators from the same community or institution that come together to collect data and generate abusive material of their peers. The harms were pervasive---victims, many of them students, were found in over 500 schools across the country, including underage victims from middle and high schools, as well as teachers~\cite{mackenzie2024deepfake}. The incident has since been referred to as the \textit{2024 South Korean Telegram Deepfake Scandals}, one of the most significant instances of organized DSV and nonconsensual image generation and distribution.

Yet while harrowing, this was far from South Korea’s first encounter with widespread, systematic DSV. Only a few years prior in 2019, a network of Telegram chatrooms---dubbed the ``Nth rooms''---had been uncovered, targeting vulnerable women and extorting them to generate humiliating and oftentimes violent sexual content to the outrage of the general public~\cite{joohee2021nth, wan_2020}. Protests and rallies publicizing the pervasive harm of spycam crimes, adult websites that shared nonconsensual and abusive sexual content, as well as the lenient responses of law enforcement, garnered increasing public support and engagement ever since their inception in 2018~\cite{shin2021beyond, minyoung_moon_digital_2024}. However, despite continued efforts in legislation, education, and grassroots activism, DSV still remains a deep-rooted aspect of South Korean society.

Why does this problem persist? What social, cultural, and technological factors sustain large-scale sexual violence across successive waves of technological change? To address these questions, we provide a sociotechnical genealogy~\cite{bardzell2020join} of DSV in South Korea. Rather than treating each incident as isolated, we trace how past formations of harm are ``recuperated, reanimated, and recombined in the present''~\cite{lin2021techniques}, examining how entrenched sociocultural misogyny is reconfigured and amplified through evolving technologies, alongside shifting legislative measures and their limitations. A genealogical approach allows us to reveal how cultures, technologies, legal frameworks, and collective practices co-constitute an evolving sociotechnical phenomenon. 

Our analysis identifies four major eras of networked DSV: (1) the pre-and early-2000s, when the intimate imagery circulation emerged through PC communication networks; (2) the rise of spycam filming in public spaces and file-sharing platforms like Soranet in the mid-2000s to mid-2010s; (3) organized sextorition networks through encrypted messaging services and social media in the late 2010s; and (4) the ongoing era of deepfake exploitation driven by generative AI.  

Through a genealogical analysis of DSV in South Korea, we identify three interrelated dimensions that shape DSV as a sociotechnical phenomenon and reveal how harm is produced, sustained, and reconfigured over time. First, we argue that “obscenity” in DSV does not pre-exist as an inherent property of images, but is actively constructed through collective sociotechnical practices that amplify participation and engagement within male-dominated online communities. Second, we demonstrate how DSV has become increasingly imperceptible, as overlapping technological systems and diversified perpetrator practices foreclose victims’ ability to recognize, anticipate, or contest harm. Third, we trace the industrialization of image-based DSV through decentralized and rhizomatic economic and criminal networks, in which women’s bodies and sexualized images are rendered as commodities and monetized across distributed infrastructures. Importantly, these dimensions are not linear developments but rather layered formations that persist, transform, and recombine across eras.  

We aim to contribute to HCI and CSCW scholarship by offering a historically grounded, culturally situated account of DSV that foregrounds the specificity of the South Korean context while generating broader insights.  
Building on the constitutive dimensions identified in our analysis, we articulate implications for future CSCW research. Methodologically, we demonstrate the value of genealogy for HCI and CSCW: rather than centering specific technologies or isolated cases, a genealogical approach enables the analysis of DSV as both rupture-introduced by emerging technological configurations-and reconfiguration of enduring dynamics that persist across periods of disuse, adaptation, and revival.




\section{Related Work}

In this section, we explore CSCW and social computing scholarship on DSV, both in terms of general technologically-mediated sexual violence as well as a specific focus on image-based sexual abuse. We also explore how cultures of misogyny, gender conflicts, and gender-based violence manifest in South Korea, providing the overall historical context as a backdrop of analysis.

\subsection{Digital and Technologically-facilitated Sexual Violence}

Digital sexual violence (DSV) refers to the various ways in which digital communication technologies are used to perpetrate criminal, civil or otherwise harmful sexually aggressive behaviors~\cite{powell2017sexual}. A large body of HCI, CSCW and privacy research explore how technology may be abused to cause harm in close, intimate partner relationships~\cite{freed2017digital, bellini_mapping_2019, tseng_care_2022, tseng2020tools, bellini2023paying}. With the added physical proximity, mobile devices, shared data, location services, as well as physical and emotional proximity are exploited by abusers to exert control over and abuse the victim, through methods such as stalking, coercive control, and even financial harms~\cite{southworth2007intimate, freed2017digital, bellini2023paying}. The sharing of sexual content between intimate partners poses additional risks, introducing data management challenges after a relationship ends~\cite{coduto2024delete} where victims could become targets of image-based sexual abuse or `revenge pornography’ ~\cite{franco2024characterizing, blancaflor2024deepfake}.

The increased prevalence and connectivity enabled through digital media provides unique affordances for gender-based violence against women in online spaces. Women and marginalized groups are often more vulnerable to violence online~\cite{wang2024counting, im2021yes, foriest2024cross, chen2022trauma}. Such violence is often networked~\cite{bhimdiwala2024fighting}, with multiple perpetrators gathering or coordinating together to harass victims ~\cite{marwick2018drinking}, and with higher likelihood of dissemination and distribution, which poses a higher risk due to the permanence of digital content.
Sambasivan et al. refers to online gender-based violence as a consequence of internet use, intended to reduce the online participation of women in an already gender-unequal Internet context~\cite{sambasivan_they_2019}. 
Thus, harassment and violence is utilized as a method of suppression, further decreasing the voice of women in these spaces. Expression of feminist or gender-critical topics often cause for women to be targeted by harassment and abuse~\cite{qin2024dismantling}, and youth are also increasingly exposed to gender-based risks~\cite{freed_understanding_2023}, such as sexual harassment, or sexual solicitations~\cite{alsoubai_friends_2022}. 

As such, it is crucial to understand technology-facilitated abuse as not just a technological issue, but as a socio-technical and a cultural issue. Women’s experiences with technology differ across cultures, as well as by demographic factors such as age, socioeconomic status, education level, etc.~\cite{powell2017sexual, batool_expanding_2024} which influences how DSV manifests~\cite{umbach_non-consensual_2024}. In Asian cultures, patriarchal social norms, institutional discrimination against women, and collectivist cultural characteristics such as family order impact how violence against women is conceptualized ~\cite{batool_expanding_2024,bansal2024scoping,umbach_non-consensual_2024,sheikh2024technology}.
We emphasize the importance of a culturally situated approach to analyzing DSV, urging the critical need to re-examine what constitutes violence and how they are normalized in cultural settings.

\subsection{Image-based Sexual Abuse}
Image-based sexual abuse is a specific type of DSV focused on the generation and sharing of sexual or sexually explicit images ~\cite{henry2015embodied,henry2018technology}. This includes dissemination of sexual visual contents that have been shared between or generated by intimate partners, as well as synthetic materials such as photoshopped images or deepfakes ~\cite{umbach_non-consensual_2024, blancaflor2024deepfake}, to threaten or extort the victim for money or sexual services (or sextortion)~\cite{powell2017sexual}. A 2019 study estimated that up to 1 out of 10 individuals participate in similar behaviors~\cite{powell_image-based_2019}, pointing to the widespread nature of such harms. 

The recent advancement and proliferation of generative AI technologies makes it easier than ever to generate sexually explicit material of an individual without their consent~\cite{umbach_non-consensual_2024, blancaflor2024deepfake, chadha2021deepfake}. 
Deepfake-generated content of high-profile individuals such as celebrities and politicians~\cite{maddocks_deepfake_2020} have brought attention to the reputational and informational harms, such as impersonation and defamation, that can arise from such technologies~\cite{bailey2021ai, freed_understanding_2023}. With an overwhelming percentage of female victims~\cite{bailey2021ai}, more than half of which are South Korean women~\cite{ajder2019state}, feminist scholars point to deepfakes as a method of circumventing consent over women's bodies, transforming them to mere sexual objects rather than individuals with agency~\cite{wagner2019word}.

Qiwei et al. explores this through a sociotechnical stack, mapping social impacts of nonconsensual image generation to its technological components such as user interfaces and hardware components~\cite{qiwei2024sociotechnical}. We build upon this perspective and suggest that there should also be a cultural element added to the stack. For example, the extent of harms caused by nonconsensually generated images, or how individuals or society respond to it, may differ by culture~\cite{shahid2022matches, batool_expanding_2024, umbach2025prevalence}. 
To this end, we explore the underlying cultural contexts through the history of image-based sexual abuse in South Korea, and aim to provide a more situated analysis of the trends and patterns observed. Specifically, we go beyond examining how harms are defined or perceived, but also how collective cultures spanning abusers, lackluster responses, and online norms contribute to normalization of behaviors surrounding image-based sexual abuse.

\subsection{Misogyny and Gender-based Violence in South Korea}

\begin{quote}
    [South Korea is...] a unique case wherein a well-run nation that has achieved great economic, technological, and political advances has seen its patriarchal values changing at a surprisingly slow pace, challenging the widely held belief that women’s overall status in society tends to improve in tandem with such progresses.
     - Hawon Jung, \textit{Flowers of Fire}  \cite{jung_flowers_2023}
\end{quote}

Oppression and violence toward women in South Korea have a deep-running history. Ever since the advent of the Glass Ceiling Index by the Economist Group in 2013, South Korea has consistently come last among OECD countries until 2024~\cite{noauthor_economist_nodate}, pointing to the persistent systematic disadvantages towards women. 
Gender-based and sexual violence also remains a critical issue in South Korea. Strict hierarchical and Confucianist cultural norms that make it hard for victims of rape or coercion-based to seek help~\cite{park2024injured, erdenebaatar2025cultural}. Physical violence such as assault, murder, and threats of violence specifically targeted towards women still remain, such as the 2016 Gangnam Station murder that kickstarted the new-wave feminism movement~\cite{lee2016feminism}.  

Women are also particularly vulnerable to various forms of online violence. In 2019, 43\% of women living in Seoul reported direct and indirect exposure to DSV, with nonconsensual image sharing emerging as the most prevalent form~\cite{seoul2019platform,kimsoojeong2024technology}. A crime analysis in 2016 by the Korean police revealed a dramatic surge in camera-based sexual crimes, rising from 3.6\% in 2006 to 24.9\% in 2015 ~\cite{kimsoojeong2024technology}. Hate speech towards women are also common, with an 2021 survey from the Human Rights Commission of South Korea reporting that the most common targets of online hate speech were women and feminists~\cite{survey2021}. In a 2018 survey, 97\% of South Korean women reported that they have been exposed to misogynistic expressions online~\cite{survey2018}. Furthermore, the massive misogynistic backlash against the feminist movement~\cite{kimjinsook2018misogyny, shin2021beyond, woo2019masculinity}, driven predominantly by male-centric online communities such as \textit{DC Inside} and \textit{FM Korea}~\cite{sun-young_kim_phenomenon_2023,sooah_kim_online_2022}, has further exacerbated the issue.

This evidence of violence toward women in South Korea is especially concerning given the rampant conservatism displayed by the male youth.~\cite{sisain2019} 
Challenges to hegemonic masculinity driven by the new-wave feminism discourse led to further backlash, with campaigns focused on misogynistic and oppressive policies toward women to utilize the antifeminist sentiment~\cite{kim2018wedisk}. 
78.9\% of men in their twenties in 2019 believed that feminism supports female supremacy rather than gender equality~\cite{sisain2019}. Existence of government organizations supporting women and gender equality, such as the Ministry of Gender Equality and Family, are often considered to be discriminatory towards men~\cite{sisain2019}, with multiple conservative political campaigns gaining popularity by promising to shut down such organizations~\cite{minyoung_moon_digital_2024}. In the 20th presidential election, the conservative candidate Yoon Suk-Yeol gathered widespread support from male voters through such promises, and was elected to office, serving as president before his impeachment in December 2024~\cite{arnold2024statistical}.

Our analysis challenges this narrative of backlash as we observe how cultural, legal, and societal responses to previous cases have enabled more recent and advanced DSV incidents. We build upon the perspective that sexual violence is not merely a consequence, but also a cause of gender inequality~\cite{armstrong2018silence,park2024injured, kim2018carceral}. As such, we explore the social and technological forces that have shaped the prevalent culture of DSV in South Korea through a genealogical approach~\cite{bardzell2020join}.



\section{Methodology}
We examine the social, cultural, and technological factors that shape recurring patterns of DSV in South Korea. Our analysis focuses on various forms of image-based sexual abuse, including nonconsensual filming, creation or manipulation of intimate imagery, and their dissemination. To ground this work, we draw on three types of sources: legal documents, news and media reports, and prior academic research. Following the integrative literature review method~\cite{whittemore2005integrative}, we move beyond the narrower scope of systematic reviews, which often restrict their attention to academic publications. Instead, we could examine diverse forms of materials, ranging from research articles and books to local legal and media texts, thereby offering a broader foundation for understanding this issue~\cite{snyder2019literature}. This allowed us to critically engage and synthesize materials that retain situated histories and epistemologies, without falling into excessive labor of translation or searching for English-language sources only. This offered us a broader and nuanced foundation for understanding this issue, leveraging the value of situated knowledge that are relatively less visible in global academic discourse \cite{parkStuckTranslationReflexive2025a}.

We adopt a historical lens oriented toward surfacing recurring patterns and dynamics
over time through systematic, contextualized interpretation~\cite{mahoney_comparative_2003, bardzell2020join}. 
Specifically, we employ a genealogical approach, which highlights how past formations are recuperated, reanimated, and recombined in the present~\cite[p.3]{lin2021techniques}. Rather than portraying the past as a linear progression into the present, genealogy foregrounds the coexistence of multiple, overlapping forms of power and control. This orientation underscores why it is important to study DSV historically not as a smooth trajectory but as layered formations in which technologies, institutions, and cultural practices continually reconfigure one another.

In line with this orientation, we frame our approach through a sociotechnical genealogy, providing a ``time-bound, critical-empirical account'' of the phenomenon~\cite{bardzell2020join}. This perspective emphasizes not only describing the sociotechnical phenomenon but also analyzing the dynamics of power that shape and regulate it~\cite{lin2021techniques}. Our approach is in line with prior HCI/CSCW scholarship that emphasizes the importance of temporal and historical analysis for understanding sociotechnical arrangements. We identify four major eras of networked, large-scale DSV in South Korea: (1) the \textbf{pre- and early-2000s}, when early networked technologies introduced the first instances of DSV, (2) the rise of \textbf{spycam and nonconsensual filming} in the early 2010s, facilitated by platforms such as Soranet and the spread of \textit{Molka}, (3) the \textbf{Nth room scandals} in the late-2010s, exposing organized sexual violence circles on Telegram, and finally (4) the ongoing era of \textbf{deepfake and peer humiliation} driven by generative AI technologies. Each era is defined by the prevailing forms of violence, the methods of perpetration, and the technologies through which they were enacted.

\subsection{Data Collection}
Our data collection proceeded in three stages: academic sources, media coverage, and legal/policy documents. The value of drawing on such a wide range of materials lies in `polymorphous engagement~\cite{gusterson1997studying}', an ethnographic orientation that highlights the importance of attending to dispersed sites and formats from everyday conversations to newspapers and official records in order to capture lived experience more fully. Similarly, our approach sought to assemble insights from both formal and popular accounts. Guided by this methodological orientation, we aimed to situate DSV incidents not only within academic debates but also in relation to media narratives, legal frameworks, and broader cultural responses.

As a first step, we conducted an extensive search of existing \textbf{academic work} addressing DSV that occurred in Korea. We used Google Scholar and DBpia\footnote{An academic information portal specialized for Korean academic materials (https://www.dbpia.co.kr/)} to search for general keywords (e.g. `digital sexual violence') as well as case-specific keywords (e.g. `Telegram', `\textit{molka}' `Nth room', `deepfake') to examine how each case is being reviewed in the literature. In doing so, we were able to review 47 relevant papers across the fields of media studies, sociology, criminology, feminist studies, and law reviews. We synthesized the findings from this work to comprehensively contextualize these crimes within the broader phenomenon of technology-facilitated abuse and misogyny in Korea.

However, we note that there is a lack of existing academic research that engages with subjects of DSV beyond legal and criminological perspectives, especially for more recent incidents. To resolve this gap, we synthesize the factual and analytical reports on incidents from \textbf{news and media coverage}, including public responses. 
The authors collected investigative news articles from both online and offline sources, focusing on established South Korean media outlets and publishers such as Hankyoreh, JTBC, and Hankook Ilbo. For the Deepfake and Nth room scandals, data collection process was guided by collaboratively constructed resources such as the ‘너머n (Beyond N)’ project\footnote{A collection of DSV-related news articles developed by the news company Hankyeoreh (https://stopn.hani.co.kr/)} and the deepfake scandal timeline by ‘빠띠 (Ppatti)\footnote{A collective resource developed by 15 Korean digital feminism activists (https://campaigns.do/tags/디지털성범죄?page=3)}'. We also conducted targeted searches using specific Korean keywords, including ``디지털 성범죄'' (digital sexual violence), ``딥페이크 성폭력'' (deepfake violence), ``딥페이크 범죄'' (deepfake crimes), as well as case-specific terminology such as ``소라넷'' (Soranet), ``N번방'' (nth room), or ``지인능욕'' (peer humiliation). To ensure the depth of information, we focused on investigative journalism pieces, with in-depth analyses of the incidents, including interviews with survivors, stakeholders, and police officers involved. We included 59 articles in our final review.

Finally, to examine how the legal and political system perceived and responded to the presented cases, we referred to \textbf{legal documents} such as legislative records and court cases. We examined legal changes that occurred as a direct or indirect result of these incidents becoming public,  reviewed existing analyses of how various digital and physical sexual violence incidents have shapend South Korean legislation. 
We examined the following four laws and their amendment histories to contextualize the amendment processes : Act on Special Cases Concerning the Punishment etc. of Sexual Crimes\footnote{성폭력범죄의 처벌 등에 관한 특례; 성폭력처벌법}, the Act on the Protection of Juveniles Against Sex Offenses (later amended and renamed as the Act on the Protection of Children and Youth Against Sex Offenses\footnote{아동 청소년의 성보호에 관한 법률; 청소년성보호법}). In addition, we referenced the Act on Promotion of Information and Communications Network Utilization and Information Protection\footnote{정보통신망 이용촉진 및 정보보호 등에 관한 법률; 정보통신망법} and the Telecommunications Business Act\footnote{전기통신사업법} to review the impact in laws governing technology. All laws were referenced from the Korean Law Information Center\footnote{https://law.go.kr/}, provided by the Ministry of Government Legislation in South Korea.

\subsection{Analysis}

We analyzed our materials through a multi-stage, iterative process as follows. First, we developed a timeline of major DSV incidents, based on thematic analysis of journalistic sources and literature. For each incident, we organized the types of technologies used to mediate and facilitate DSV, while identifying the repercussions, social and legal responses to each incident. During these individual case readings, we attended to the specificities within the Korean context, including: 1) salient cultural characteristics, 2) the ways each case aligned with patterns of misogyny prevalent in Korean digital/offline cultures, 3) inherent technological affordances that shaped the operations of DSV, and 4) sociocultural factors that influenced public perception and interpretation of victim's experiences.

Following that, we explored relationships and recurring patterns across cases to identify larger patterns and trends in DSV. Our analysis was guided by the following analytical questions: (1) the impact of the deployed technologies in facilitating DSV, (2) how women were exploited through these technologies, (3) how different stakeholders became involved, (4) how these dynamics were shaped by South Korea’s sociocultural and institutional context, and (5) what particular problems or challenges were identified in individual documents.

Between October 2024 and May 2025, the six co-authors held weekly discussions in which we continually refined interpretations and deepened insights. 
These sustained conversations moved us beyond incident level analysis toward comparative understanding and reasoning, enabling us to surface both micro-level, interpersonal mechanisms and macro-level structural and systematic implications. To deepen and challenge our interpretations, we reorganized the full set of incidents and cases through four analytic lenses drawn from the team's diverse expertise, namely (1) the `fakeness' and manipulability of digital artifacts; (2) community-driven modes of abusive content production and circulation; (3) the dual role of technology as an enabler of harm and a tool for detection; and (4) attitude and behaviors of individuals who consume or redistribute abusive materials. We also revisited conceptual frameworks from prior research on DSV (e.g., \cite{maddocks2020deepfake, gosse2020politics}) to situate our readings, identify blind spots, and strengthen our findings with established analytic constructs.






Through these iterative conversations, we were able to interrogate and cross-check our interpretations and assumptions in relation to our positions, and surface new themes.
Crucially, we emphasize that our interpretations were situated on our values and life trajectories. Taking a genealogical analytical process, we foregrounded our lived experiences as South Korean feminist researchers to make informed judgments of the materials. We reflect on our positionality in the following section. 


\subsection{Positionality of Research Team} \label{sec:positionality}

The research team is composed of an interdisciplinary group of social computing, HCI and interaction design researchers, all of whom are native South Koreans and identify as women. The team’s diverse expertise across various domains of HCI research-including ethics, labor, embodied technology, feminist/Queer HCI, Human-AI interaction, social computing, AI policy, digital harms, and violence-enable a multi-faceted examination of the issue of DSV in South Korea. Having directly experienced and witnessed the impacts of misogyny and gender-based violence perpetuated through technology, our analysis is deeply informed by our personal and shared experiences of growing up, living, and being educated in South Korean society as women and/or as intersectional feminists \cite{crenshawMappingMarginsIntersectionality1991}. These lived experiences within the cultural, linguistic, and institutional context allowed us to read against the grain of official narratives, to detect silences or implicit meanings in media accounts, and to situate individual cases within broader trajectories of South Korean gender politics. These experiences shape our positionality and analysis, while providing critical contexts for the research we present. Rather than treating interpretation as a purely technical exercise, we foreground our shared background and experiential knowledge as an essential lens for making sense of this phenomenon.

\section{Four Eras of DSV: A Case Study of Image-based Sexual Abuse in South Korea} \label{sec:casestudy}

In this section, we present four distinct eras of imaged-based DSV in South Korea. Each era was divided and conceptualized by dominant technologies being in use to create and share nonconsensual and/or sexualized imagery. We present prominent cases, narrating their emergence and evolution in conjunction with technological advancements. Our periodization is not intended to suggest a linear progression of harm nor present exhaustive accounts of criminal abusive practices that took place. Instead, we present it as an analytical framework to elucidate how forms and characteristics of image-based DSV evolve in relation to shifting technological affordances, legislative measures, and persisting culture of misogyny. A summary of relevant technological factors and corresponding legislative changes can be found in Table \ref{tab:geneaology} \footnote{For genealogical analysis, we map each era with representative crime cases and main technologies being in use. However, we acknowledge that perpetrating patterns and technologies that were used in earlier era often carried on in the latter era.}

\begin{table}
\centerline{}
\scriptsize
\begin{sideways}
\begin{tabular}{>{\raggedright\arraybackslash}p{0.1\linewidth}|>{\raggedright\arraybackslash}p{0.2\linewidth}|>{\raggedright\arraybackslash}p{0.2\linewidth}|>{\raggedright\arraybackslash}p{0.7\linewidth}}
\toprule
\textbf{Era} & \textbf{Image Production Technology} & \textbf{Sharing Technology} &  \textbf{Corresponding Legislative Change} \\
\midrule
Pre- and Early- 2000s & Portable recording technology (e.g. Digital cameras, Camcorders) & Peer-to-peer sharing, PC networks,  internet technology & 
{\textbf{\textless{}Act on the Punishment of Sexual crimes and Protection of Victims\textgreater{} (1994)} \newline
\vspace{-9pt}
\begin{itemize}[itemsep=0pt, leftmargin=2em]
    \item Banned digital sexual harassment and its delivery (1994)
    \item Banned nonconsensual \textit{filming} of others in \textit{sexual situations} (1998)
    \item Specified increased statutory sentence and additional punishment for delivery of non-consensual filming crimes (2006)
    \item Banned the \textit{selling, screening,} and \textit{distributing}  of non-consensually filmed materials (2006)
\end{itemize} 
\vspace{1.5pt}
\textbf{\textless{}Act on the Protection of Juveniles Against Sex Offenses\textgreater{} (2000)}\newline
\vspace{-9pt}
\begin{itemize}[itemsep=0pt, leftmargin=2em]
    \item Banned \textit{production} and \textit{distribution} of juveniles pornography (2000)\vspace{2.5pt}
\end{itemize} 
} 
\\
\midrule
Mid 2000s to Mid 2010s & Portable, subminiature recording devices (e.g. Digital cameras, Camcorders, Smartphones) & P2P file sharing / Large-scale storage services (Webhard), Streaming services & {
\textbf{\textless{}Telecommunications Business Act\textgreater{}} \newline
\vspace{-9pt}
\begin{itemize}[itemsep=0pt, leftmargin=2em]
    \item Introduced \textit{online service providers' responsibility} to moderate for and ban nonconsensually filmed content on their services (2018)
\end{itemize} 
}\\
\midrule
Late 2010s & Personal data (e.g., Photos posted on social media) & End-to-end encrypted messaging services, networked social media platforms & {\textbf{\textless{}Act on the Protection of Children and Youth Against Sex Offenses
\textgreater{}}\newline
\vspace{-9pt}
\begin{itemize}[itemsep=0pt, leftmargin=2em]
    \item Specified increased statutory sentence and additional punishment for sexual violence crimes targeting children and youth (2020)
    \item Article preventing children and youth victims of sexual trafficking to be punished as perpetrators introduced (2020)
    \item Changed the term “\textit{obscene} materials involving children and adolescents” to “child and adolescent \textit{sexual exploitation} materials.”(2020)
\end{itemize} 
\vspace{1.5pt}
\textbf{\textless{}Act on Special Cases Concerning the Punishment etc. of Sexual Crimes\textgreater{}}\newline
\vspace{-9pt}
\begin{itemize}[itemsep=0pt, leftmargin=2em]
    \item Specified the banning of \textit{fake or edited} media content for sexually harassing purposes (2020)
    \item Specified increased statutory sentence and additional punishment for digital sexual violence and nonconsensual filming and its distribution (2020, 2024)
    \item Banned the \textit{owning, purchasing,} and \textit{consuming} of nonconsensually filmed content, as well as blackmailing of individuals through sexual materials (2020)
    \item Banned \textit{nonconsensual distribution} of sexual images, regardless of \textit{consent to filming} (2020) \vspace{2.5pt}
\end{itemize} 
} \\
\midrule
Late 2010s to Early 2020s & Generative-AI based photo/video editing technology & End-to-end encrypted messaging services & {\textbf{\textless{}Act on Special Cases Concerning the Punishment etc. of Sexual Crimes\textgreater{}} \newline
\vspace{-9pt}
\begin{itemize}[itemsep=0pt, leftmargin=2em]
    \item Clarified that there \textit{does not need to be intent to distribute} for owners of nonconsensually shared sexual content to be persecuted (2024)
    \item Banned the owning, purchasing, and consuming of nonconsensually \textit{generated synthetic} content (2024)
\end{itemize} 
\vspace{1.5pt}
\textbf{\textless{}Act on the Protection of Children and Youth Against Sex Offenses
\textgreater{}}\newline
\vspace{-9pt}
\begin{itemize}[itemsep=0pt, leftmargin=2em]
    \item Requires judicial police officers to request deletion or access blocking of child sexual exploitation material through the Korea Communications Standards Commission (2024)
\end{itemize}
\vspace{1.5pt}
\textbf{\textless{}Act on Promotion of Information and Communications Network Utilization and Information Protection
\textgreater{}}\newline
\vspace{-9pt}
\begin{itemize}[itemsep=0pt, leftmargin=2em]
    \item Article urging the Ministry of Science and the Korea Communications Commission to establish legislative measures to prevent harm from using AI technology to nonconsensually generate audio or video content of individuals (2023) \vspace{2.5pt}
\end{itemize}}
\\
\bottomrule
\end{tabular}
\end{sideways}
\normalsize
\caption{Technological factors involved in and major legislative changes introduced as a result of each era of digital sexual violence crimes in South Korea. This table does not include an exhaustive list of all legislative changes.~\cite{sooah2024, jeon2021n}}
    \label{tab:geneaology}
\end{table}

\subsection{Era 1. Pre- and Early 2000s: The Emergence of Intimate Imagery Circulation through PC Communication Networks} \label{sec:casestudy-1}
The late 1990s and early 2000s marked the initial emergence of digitally mediated intimate imagery circulation in South Korea, enabled by early internet infrastructures, personal computers, and anonymous online communication. South Korea was one of the first countries to widely adopt Internet infrastructures in 1982, followed by rapid advances in information technology and connectivity that led to Korea becoming one of the leading IT presences worldwide~\cite{chon2013history, chon2005brief}. These were facilitated through locally developed PC network platforms such as Cheonrian and Hi-Tel~\cite{kim2008internetkorea, cheonrian2000}.

However, this period of technological advancement encompassed the introduction of DSV, starting in the late 1990s~\cite{jeon2021n}. In this period, portable recording devices such as cameras and camcorders were used to film, copy, and distribute nonconsensual imagery. Recorded cases of such crimes date as early as 1997, when high school students nonconsensually filmed videos sexually assaulting a female middle school student in the infamous \textit{Red muffler incident}~\cite{sorakim2018changes}. The videotape, and others like them, were copied and shared first through peer groups such as school communities, but soon escalated and spread to commercial vendors such as the video shops at Sewoon marketplace~\cite{muffler1997}.

While these videos were constrained to physical videotapes, online platforms served as a site for proliferation and distribution. `Adult-only' boards of PC network platforms such as \textit{Cheonrian} saw users asking to purchase the \textit{Red Muffler} video~\cite{muffler1997}.
Online chat rooms and anonymous forums facilitated sexual exploitation, including the sharing of information related to statutory rape and sexual trafficking~\cite{jeon2021n}. As a virtual space harnessing networked anonymity, these platforms exposed women and minors to persistent sexual harassment and abuse enabling the rapid and extensive spread of nonconsensual sexual abuse materials.

South Korea officially criminalized sexual violence and assault in 1994, with the introduction of the \textit{Act on the Punishment of Sexual Crimes and Protection of Victims}. Notably, the Act addressed what would later be understood as “digital” sexual violence from its inception. It included an article about sexual violence in mediated spaces, where the use of communication media to sexually harass others were considered as a punishable offense. Soon after, it was amended in 1998 to additionally criminalize acts of nonconsensual filming of others in sexual situation, as well as the selling, screening, and distribution of such content in 2006. Acknowledging its limitations in separately addressing the punishment of perpetrators and the support of victims, this law was abolished and its legal action was divided into the \textit{Act on Specal Cases Concerning the Punishment etc. of Sexual Crimes} and the \textit{Act on the Prevention of Sexual Assault and Protection, etc. of Victims Thereof} in 2020. 

Similarly, the \textit{Act on the Protection of Juveniles Against Sex Offenses} was introduced in 2000, specifying the production and distribution of juvenile pornography as illegal. This law was also later amended to the \textit{Act on the Protection of Children and Youth Against Sex Offenses} to include children in the legal protection boundary.
However, these legal measures failed to provide adequate protections for victims. The victim of the \textit{Red muffler} incident was arrested and was sentenced to probation over charges of creating obscene materials, despite being a victim of physical and digital sexual violence ~\cite{muffler1997,muffler1997sentence}. This case exposes the law’s limitations in accounting for contextual factors, revealing a structural tension between the punishment of perpetrators and the protection of victims.
As a result, cultures of DSV still persisted, laying out the foundation for an evolving culture of digital sexual abuse and setting out a long history of pervasive DSV in South Korea~\cite{sooah2024}.

\subsection{Era 2. Mid 2000s to Mid 2010s: Spycam Filming in Public Spaces and Distribution through File-Sharing Platforms} \label{sec:casestudy-2}

Moving into the mid 2000s, high-profile incidents of sexual violence towards minors and children further exposed structural misogyny in South Korea. This prompted later reforms such as 2012 amendment banning the possession and consumption of child sexual abuse material~\cite{sooah2024, jeon2021n}.
During this period, a new form of abuse gained traction, driven by the convergence of portable and subminiature recording devices and large-scale file-sharing platforms. Nonconsensual images taken in (semi-)public spaces, dubbed as \textit{molka} (A portmanteau denoting `secret filming' in Korean), circulated through profit-driven online platforms, known as Webhards\footnote{A jargon used in South Korea to denote Storage as a Service(STaaS)}, such as WeDisk~\cite{kim2018wedisk}. 

Public attention intensified following the 2015 Waterpark Molka Crimes, which revealed the organized production and commercial circulation of nonconsensual spycam footage taken from recreational facilities~\cite{soranet_ban}. A male perpetrator recruited a female accomplice and instructed her to secretly film in women’s shower rooms across water parks and outdoor swimming pools~\cite{MBN_2015}, for a compensation of approximately 2,000,000 KRW (about USD \$1,500)~\cite{etoday_2015}. Police investigations later confirmed that the accomplice had produced approximately 185 minutes of illicit footage~\cite{etoday_2015}. The perpetrator monetized this material through multiple channels, including selling portions of the videos via an internet messenger service for 1,200,000 KRW(about USD 800)~\cite{MBN_2015}. Additional dissemination occurred through adult websites hosted on overseas servers. Similar incidents were subsequently reported in other communal spaces, including public restrooms and public transportation systems.

\subsubsection{Soranet and Participatory Cultures of Consumption}
The Soranet case is an archetypal form of DSV in this era. Production, sharing, and collective consumption of nonconsensual intimate imagery and derogatory sexually objectifying textual content of women's bodies became normalized through homosocial online community participation.  
Originally launched as an adult content website and ever since its renewal in 2003 as a membership-based portal site ~\cite{kim2023n}, Soranet quickly became a place where its members uploaded, labeled, and appraised nonconsensually filmed image/videos. Access to full participation required users to register as ``creators,'' a status granted through the submission of voyeuristic images such as spycam footage targeting women's bodies~\cite{gobang}. Advancement of status depended on producing content of increasing severity, with higher-ranking creators gaining recognition and status among members~\cite{mtoday_soranet_2015}.

Beyond a repository of illicit content, Soranet also served as a site for real-time coordination of sexual violence. The platform regularly hosted posts soliciting participants for sexual assault, a practice colloquially referred to as ``chodaenam (초대남)'', recruiting men to participate in group sexual assaults~\cite{mtoday_soranet_2015}. Once invited, participants would sexually assault women incapacitated by severe intoxication and document their participation by inscribing their online pseudonyms on the victim’s body and submitting photos for verification. These images operated simultaneously as evidence of compliance and as symbolic tokens of status within the community. In addition, Soranet frequently hosted forms of revenge pornography, including the intentional disclosure of intimate images of acquaintances or former partners, often accompanied by personally identifying information.

Within this environment, online misogyny was (re)produced through the development of derogatory neologisms stereotyping Korean women \cite{kimjinsook2018misogyny}. For instance, male users commonly expressed ill-founded condemnation toward Korean women by combining certain epithet with \textit{-nyeo/nyeon}(girl, woman/bitch). Terms such as \textit{Doenjang-nyeo}(soybean paste girl) was widely used, also in mainstream media, to stigmatize certain young women as vain and materialistic, which later expanded to \textit{Kimchi-nyeo/nyeon} to contemn Korean women in general \cite{kimjinsook2018misogyny}. Users also referred to nonconsensually recorded images and videos of Korean women as \textit{Gooksan} (국산), a word denoting domestic or `Made in Korea' products ~\cite{chang_study_2018, jung_conception_nodate}. Comment sections included offensive and demeaning appraisals on victims' appearances and sexual desirability. Bolstered by anonymous relationships, Soranet naturalized sexual voyeurism and fragmentation of women's bodies. 

Given the malevolence of such content, the Korea Communications Standards Commission (KOCSC) made over 200 attempts to block access to Soranet~\cite{soranet_ban}. However, as the server was operated overseas, the administrators evaded this by continuously updating the web domain and sharing the changes to its users through social media such as Twitter \cite{mediatoday__2015}. Grassroots feminist activists gathered in online communities and social media to launch an organized campaign called Digital Sexual Violence Out (DSO). Their call for legal action served a critical role in shutting down Soranet in June 2016, after 17 years of operation and over 1 million aggregate users \cite{soranetserver_jtbc}. While the closure of this platform marked a symbolic victory, it did not dismantle the broader ecosystem of abuse that had taken its shape. Victims still found themselves unsupported by law-enforcement systems, where police would often dismiss reports of DSV by claiming that it is too hard to track down~\cite{lee2020track, jung_flowers_2023}. This later became the seed for prominent women’s rights marches and protests that began in 2018, when the public started recognizing the gender discriminatory responses of law enforcement – where male victims were taken much more seriously and their perpetrators were tracked down more quickly and efficiently, in contrast with numerous cases where the victims were women~\cite{jung_flowers_2023}. Moreover, it led to revisions of the \textit{Telecommunications Business Act}, marking a significant shift in regulatory responsibility by imposing direct obligations on online service providers, including Soranet, to moderate and prohibit the distribution of nonconsensually filmed content, rather than highly relying on regulatory oversight by the KOCSC.

\subsubsection{Webhard Cartels and the Commercialization of Nonconsensual Images} \label{case:webhard}

While platforms such as Soranet foregrounded the normalization of nonconsensual imagery consumption, the Webhard Cartels exposed the institutionalization of imagery retail ecosystem via profit-oriented infrastructures. Webhards enabled a long-term storage, re-distribution, and monetization of spycam materials long after its initial upload.

Public scrutiny of the so-called ``Webhard Cartel,'' revealed a coordinated ecosystem involving webhard platform operators, heavy uploaders, filtering companies, and digital undertaker services~\cite{webhardcartel_2025}. 
A 2018 police investigation revealed that many of these services were covertly linked to the webhard platforms hosting the content~\cite{cartel2018}. The cycle of harm was as follows~\cite{lee2023digitalpolis}: producers and resharers uploaded nonconsensual images and videos to webhards for profit; victims, seeking relief, turned to digital undertaker services to request takedowns; and these services claimed to provide filtering technologies capable of detecting illicit content across webhard databases. In practice, however, the digital undertaker and filtering companies were often owned or operated by the same individuals who controlled the webhard platforms themselves. This arrangement allowed operators to ignore reuploads of previously deleted material and to repeatedly extract fees from victims, effectively monetizing both circulation and removal. In short, the webhard platforms, digital undertakers, and their subcontractors had formulated a discreet cartel gaining profit from the (re-)distribution, filtering, and deletion of content, feeding on victims' fear and trauma. According to prosecutors, approximately 3.88 million illegal videos, including sexually exploitative materials, were linked to the operations of a central figure in the Webhard Cartel, generating profits estimated at 35 billion KRW~\cite{webhardcartel_2025}.

\subsection{Era 3. Late 2010s: Organized Sextortion Targeting Vulnerable Minors through Social Media and Encrypted Messaging Services} \label{sec:casestudy-3}

By the late 2010s, DSV in South Korea had entered a new phase marked by organized sextortion facilitated by encrypted messaging services such as Telegram. These platforms afforded perpetrators advanced anonymity through encryption while enabling the rapid circulation of illicit material. Therefore, perpetrators could both coordinate a large number of co-conspiring participants and identify, recruit, and control victims at scale while minimizing the risk of detection. 

In July 2019, Team Flame, an investigative journalism collective led by two university students, uncovered an extensive network of Telegram-based chatrooms dedicated to the circulation of sexual violence content while investigating cases of illegal filming in South Korea~\cite{tbs_nth_room_2020}. Their investigation documented the real-time distribution of sexually exploitative materials, including content involving minors, as well as coercive and abusive interactions among participants~\cite{shindonga_telegram2020}. Continued undercover monitoring ultimately led to public exposure in late 2019, with mainstream outlets collectively referring to these interconnected chatrooms as the ``N-th room cases''~\cite{hankyoreh2019telegram}.

The N-th room case illustrates how sexual exploitation networks emerged and expanded through a combination of coercion, anonymity, and platform affordances. The network originated from the systematic targeting of so-called ``deviant accounts (일탈계)'' on social media, which often featured revealing or sexually suggestive images and text~\cite{mt_telegram2022, beminor_digital_sexcrime2023, telegram_punished_2019, telegram_2019_woman}. The earliest and most systematic exploitation of these accounts was initiated by an operator known as ``Godgod (갓갓)''. He and his accomplice subjected victims to hacking, impersonation of law enforcement, and sustained blackmail. Fearing exposure and social stigma among friends, family members, or employers, victims were coerced into producing and sharing increasingly violent sexual content~\cite{telegram_punished_2019, telegram_2019_woman}. Although these actions were forced, perpetrators framed the resulting materials as evidence of voluntary participation, further entrenching victims’ vulnerability. 

Over time, the content was distributed across multiple interconnected Telegram chatrooms with differentiated access conditions, including free entry points and paid rooms that functioned as an integrated funnel for recruitment and monetization~\cite{park_hani_2022}. Free-access rooms operated alongside pay-for-access rooms, serving as entry points where sexually exploitative materials were intermittently distributed at no cost. These free rooms simultaneously functioned as advertising spaces, where administrators actively promoted as well as emphasized the exclusivity and severity of content available behind paywalls. Access to paid rooms typically required participants to transfer prepaid digital gift certificates worth approximately 20,000 KRW (about 13.5 USD), directly to the room operator~\cite{park_hani_2022}. Once admitted, participants were encouraged to make additional payments to obtain further content or privileges. Participants were organized into a hierarchical system based on accumulated “experience points,” with ranks such as “public official,” “upper class,” “citizen,” “middle class,” and “single-parent household,” reflecting a deliberate gamification of participation~\cite{jo_hani_21, nthroom_baksa}. According to prosecution records, this tiered structure generated massive financial returns, with an estimated hundreds of thousands of paying members and escalating entry fees proportional to the severity of content.

This shift toward a more centralized and profit-oriented model emerged through successive adaptations within the N-th room ecosystem. The initial infrastructure and coercive techniques developed by Godgod were later inherited and maintained by operators known as Watchman (와치맨) and Baksa (박사). The `Doctor’s Room', operated by Baksa, developed as a derivative of this earlier system, retaining its core monetization logic while substantially expanding its organizational complexity~\cite{park_hani_2022}. In this room, perpetrators recruited women through social media and messaging applications by offering `sponsorship' arrangements~\cite{baksabaksa}. The  images and personal information obtained during these interactions were subsequently used to coerce and exploit the victims into producing additional content.

\begin{figure*}[t]
   \centering
   \includegraphics[width=\textwidth]{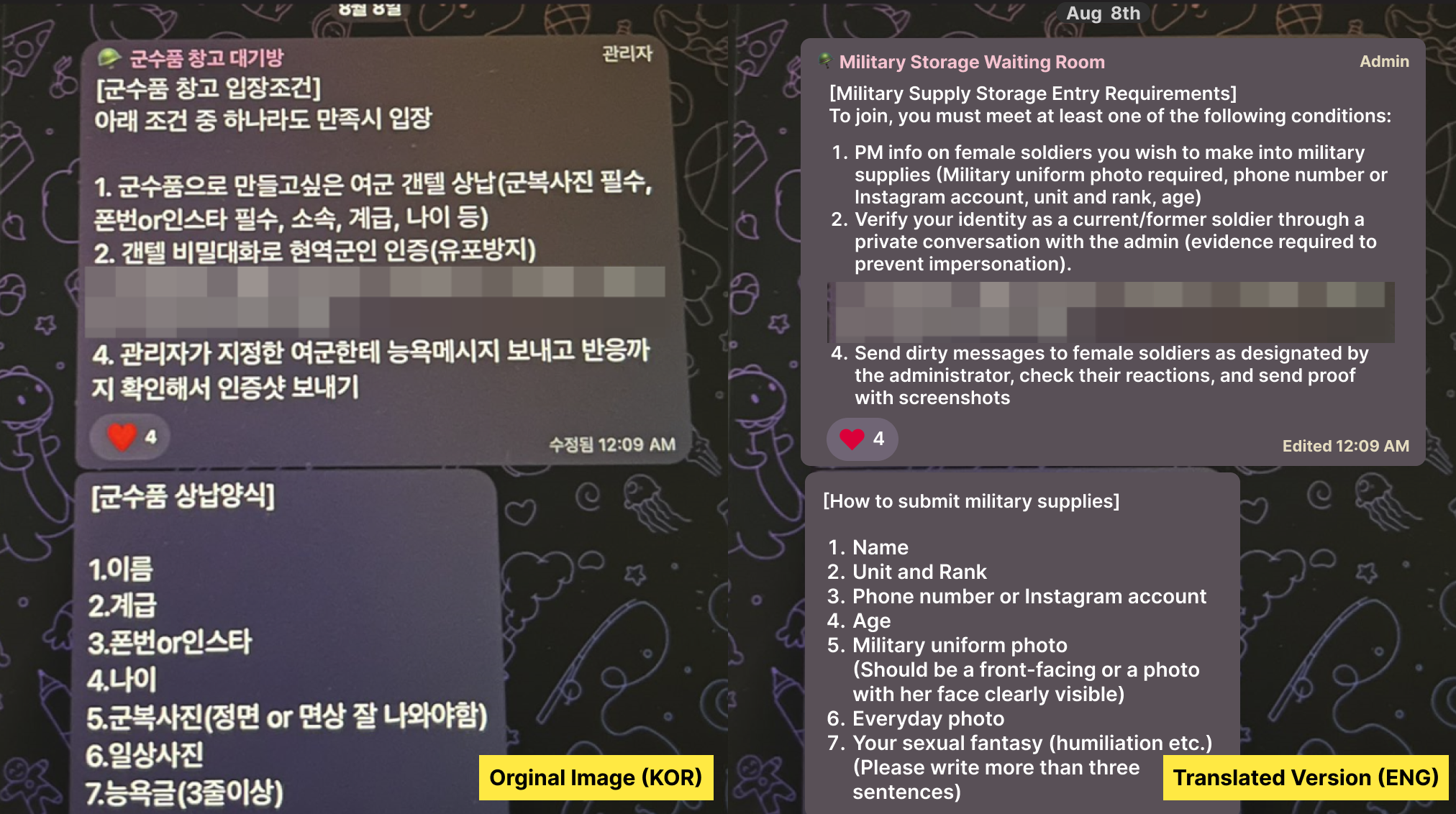}
   \caption{In a Telegram group chat suspected to include active-duty soldiers, female soldiers were referred to as ``military supplies,'' and requests for victim photos to be used in deepfake crimes were described as ``tribute forms.''}
   \label{fig:img1}
\end{figure*}


A defining characteristic of both the N-th room ecosystem and its later derivatives was the deliberate production of exclusivity through tightly controlled access. 
As abuse increasingly targeted individuals embedded within shared social or institutional contexts, identity verification became a central organizing mechanism. Many chatrooms required interviews, victim images, or documentation proving affiliation, such as graduation photos or institutional identification cards~\cite{military2024}.
One military-based chatroom (Figure~\ref{fig:img1}) exemplifies this logic through a multi-stage verification regime. Prospective members were required to submit personal military service records and provide evidence that they had sent a sexually harassing private message to an unacquainted female soldier designated by administrators as an entry requirement~\cite{military2024}. 

Although this process involved the submission of highly sensitive and personally identifiable information that could potentially be used against participants, a shared culture of conformity and mutual implication fostered a perception of safety despite the evident legal risks. Through such mechanisms, participation in abuse was reframed as adherence to community norms rather than an individual act of deviance. By complying with administrators’ rules and escalating demands, participants became progressively more embedded in the group’s antisocial and exploitative practices. In this sense, the approval and verification process functioned as a classic “foot-in-the-door” mechanism~\cite{freedman1966compliance,dillard1984sequential}, incrementally reinforcing commitment through active participation in abuse. Similar logics of exclusivity and verification have continued to shape later peer-based humiliation and deepfake exploitation cases that characterize Era 4.

This public pressure catalyzed a wave of legislative reforms commonly referred to by the media as the `Nth room prevention laws' by major news outlets. News reports and experts highlighted victims’ reluctance to seek legal assistance, due to fears of being treated as co-offenders, exposing significant gaps in existing legal frameworks~\cite{han2019dark, telegram_punished_2019}. 
Critics also pointed to the government’s delayed response to digital sexual violence and its failure to advance relevant legislation in a timely manner~\cite{noauthor_n_nodate}. Consequently, a package of legal reforms spanning criminal law, child and youth protection, and sexual violence statutes was designated as a priority task of South Korea’s 20th National Assembly \footnote{The 20th National Assembly began its tenure on the 30th of May, 2016, and closed on the 29th of May, 2020}~\cite{preventn2020}.  In particular, amendments to the \textit{Act on the Protection of Children and Youth Against Sex Offenses} were introduced to prevent children and youth from being punished as perpetrators of sexual trafficking, as well as re-framing sexual materials involving children and youth from `obscene materials' to `sexual exploitation' materials in 2020. Similarly, the \textit{Act on Special Cases Concerning the Punishment etc. of Sexual Crimes} increased penalties for digital and sexual violence offenses and introduced measures to prevent secondary victimization, such as criminalizing the possession, purchase, and consumption of nonconsensually produced sexual materials. The 2020 reforms also specified that consent to film and distribute are distinct by banning the nonconsensual distribution of sexual images regardless of whether there was consent to film.


\subsection{Era 4. Late 2010s to Mid 2020s: Circulation of Nonconsensual AI-Generated Intimate Deepfakes on Social Media} \label{sec:casestudy-4}

Following the public exposure of organized sextortion networks, DSV further diversified through the widespread availability of generative image technologies~\cite{soranetpersecution, park2024deepfake} and encrypted communication infrastructures like Telegram~\cite{military2024}. Unlike earlier forms of abuse, in which perpetrators had to make physical or digital contact with victims, harms in this era were increasingly materialized with nonconsensual synthesis of existing personal data (e.g., profile image in messaging app) into sexualized content~\cite{military2024}. Advances in image and video generative AI technologies have made it easy to produce, replicate, and circulate synthetic sexual material at scale and speed, enabling forms of abuse that no longer depend on direct contact, coercion, or even the victim’s awareness.

Building on the organizational, technological, and cultural logics consolidated in Era 3,  practices commonly referred to as `peer humiliation (지인능욕)' developed. In these cases, sexualized content is superimposed onto the victim’s identifiable personal data, such as photos, names, affiliation, residential address, and school/workplace~\cite{sorakim2018changes, beyond_21}. 
As in the earlier eras, perpetrators frequently harvested everyday images from victims’ social media accounts and repurposed them for harassment, including posting sexually explicit captions or inviting others to participate in coordinated humiliation. In many cases, images were digitally altered or superimposed onto pornographic material to degrade victims through forced sexual association. While earlier instances relied on manual photo and video editing techniques, the diffusion of generative AI technologies in the early 2020s 
accelerated and standardized these practices~\cite{choi_2024}, dramatically lowering technical barriers to participation.

These dynamics increasingly involved underage perpetrators.
In one case, a teacher discovered deepfake-generated sexual images of herself when reviewing the phone of a student who had secretly taken under-skirt photographs of other teachers~\cite{park2024student}.
In another case, a student impersonated a teacher by creating a fake social media account, posting fabricated messages alongside deepfake generated sexual images, using photographs her teacher’s wedding family photos. These materials were circulated using the teacher’s real name in social media, amplifying visibility and harm~\cite{park2024student}. Such cases highlight how emerging technical infrastructures reshaped not only how abuse was produced and distributed, but also who could participate in it.

\begin{figure*}[t]
    \centering
    \includegraphics[width=\textwidth]{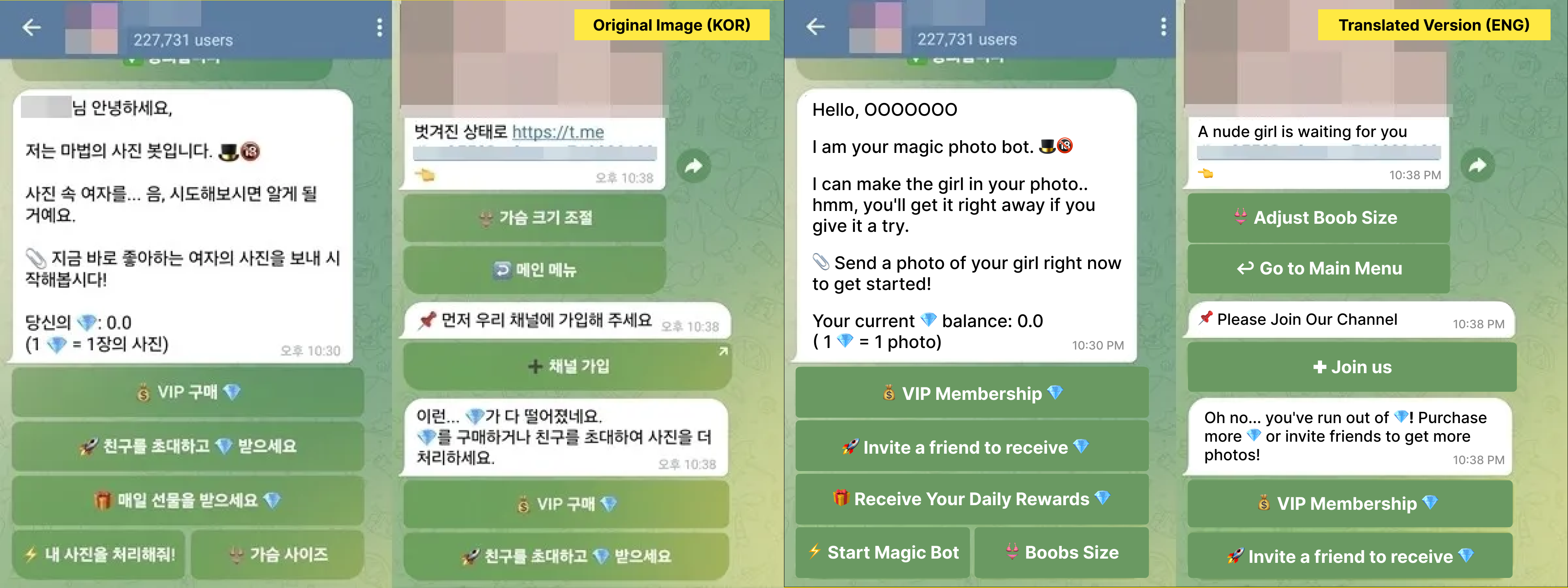}
    \caption{A Telegram room with over 220,000 participants was involved in deepfake sexual exploitation. The platform utilized an AI bot that could generate deepfake sexual content within 5 to 7 seconds when a photo of a woman was uploaded.}
    \label{fig:img2}
\end{figure*}

The scale and organization of this phenomenon became publicly visible during the 2024 South Korea deepfake scandals, when police investigations uncovered extensive Telegram networks devoted to the nonconsensual creation of sexualized images of women~\cite{park2024deepfake22} and children~\cite{military2024, nocutnews_6343786}.
Whereas earlier image-based abuse relied on editing tools such as Photoshop and individual labor~\cite{ART003188329}, deepfake technologies including `nudifying apps' automated image manipulation, allowing perpetrators to generate synthetic sexual content from ordinary photographs within seconds~\cite{military2024}. Telegram bots further automated this process by offering end-to-end pipelines (Figure \ref{fig:img2}): users could purchase in-platform currency (“diamonds”) to generate images, earn diamonds by inviting others, create explicit images using preset parameters, and refine outputs by altering physical attributes. While users were initially granted a small number of free generations, subsequent requests incurred per-image fees, typically processed through cryptocurrency to preserve anonymity. With reported prices as low as approximately 2,000 KRW (USD 1–2) per image and around 13,000 KRW (USD 10) per video, the cost of generating illegal synthetic sexual content was sufficiently minimal to lower barriers to experimentation and repeated abuse. 

These automated infrastructures blurred conventional distinctions between producers and consumers of abuse. For example, in one widely reported case from 2021~\cite{donga_deepfake2024}, a middle school student initially commissioned a deepfake sexual image of two 15-year-old girls online. After learning the production techniques himself, the student went on to create 177 deepfake images, including that of children, which he then sold through Telegram chatrooms. This case illustrates how the accessibility of deepfake technologies enables rapid transitions from consumer to producer, collapsing traditional role boundaries within the abuse economy. 

At the center of this ecosystem are operators who develop and maintain deepfake bots, automated programs that generate synthetic sexual images using AI and are distributed through Telegram channels. These bots structure production and exchange through tiered pricing models, in which higher payments grant access to more explicit, higher-resolution, or more detailed outputs~\cite{donga_deepfake2024}. Transactions are conducted through in-platform virtual currencies such as “diamonds,” which users purchase using cryptocurrency~\cite{donga_deepfake2024, military2024}, further obscuring financial trails and preserving anonymity. As demand and transaction volume increase, both bot operators and the hosting platform benefit economically, reinforcing incentives to expand and sustain these infrastructures~\cite{donga_deepfake2024}. For victims, however, this infrastructural expansion transforms harm from a discrete act into an ongoing condition of exposure and uncertainty.


Even in Era 4, these synthetic images are absorbed into opaque AI bot infrastructures~\cite{military2024}, where victims had no way of knowing whether their images were temporarily processed, permanently stored, reused for further generations, or incorporated into training data. In this sense, harm is not fully contained by legal intervention. As deepfake bots were designed as automated systems that mediate content generation without sustained person-to-person interaction, it left little identifying information about individual operators~\cite{fnnews_202409101825398415}. These limitations are compounded by Telegram’s lack of cooperation with law enforcement, which further restricts access to account-level data and conversation data. As a result, even when perpetrators can be identified and punished, victims may continue to be exposed to fear of yet-unknown future harm.

These conditions highlight the limits of case-by-case enforcement and help explain why legal and policy interventions increasingly focus on infrastructure, platforms, and enduring forms of harm. In response to the expansion of AI-leveraged sexual violence, South Korea has introduced a series of legal and policy interventions that reflect growing recognition of the infrastructural and enduring nature of harm in the generative AI era. Recognizing the role of interconnected actors within this ecosystem, revisions to the \textit{Act on the Protection of Children and Youth Against Sex Offenses} required judicial police officers to request deletion or access blocking of child sexual exploitation material through the KCSC. At the same time, changes to the \textit{Act on Promotion of Information and Communications Network Utilization and Information Protection} urged regulators to address the use of AI technologies to generate nonconsensual audio and visual content. In response to the interconnected and evolving forms of harm, amendments to the \textit{Act on Special Cases Concerning the Punishment of Sexual Crimes} clarified that the possession of nonconsensually shared sexual content constitutes an offense even without intent to distribute and explicitly banned the owning, purchasing, and consumption of nonconsensually generated synthetic sexual content.




\section{Three Dimensions of DSV: A Genealogical Analysis of Cases in South Korea} \label{sec:findings}

In this section, we present three key dimensions of DSV by taking a genealogical analysis on prominent cases presented in Section \ref{sec:casestudy}. In describing these dimensions, we deliberately leap between different eras and cases within to highlight \textit{ruptures and continuities}~\cite{jacobsen_tensions_2024} of sociocultural, technological, and legal factors that transfigure how each dimension manifests. The dimensions are as follows. 


\subsection{``Obscene'' Imagery as Oppressive Control Over Women powered by Homo-social Male Solidarity} \label{sec:findings-1}

In South Korea, the notion of obscenity has repetitively appeared in the public and legal discourse to describe and interpret image-based DSV. For instance, when the Red Muffler Incident was publicized during the first era (see section \ref{sec:casestudy-1}), mainstream journalism reported individual traits about the victim -- that she was a juvenile delinquent and a teenage runaway -- and portrayed her as one of the perpetrators who have conspired in “distribution of obscene materials (음화반포죄)” under the Criminal Act \cite{muffler1997sentence}. Two decades later, when the Nth Room case broke out, numerous media coverage highlighted and expressed apprehension that many victims were owners of “deviant accounts” on social media, posting sexual and intimate imagery of themselves \cite{beminor_digital_sexcrime2023}. The initial obscene deeds of the victims and their consequential risks were again put in the spotlight. This framing was also reflected in legislative language, where terms such as “obscene materials” were used to refer to sexually exploitative materials of children and adolescents prior to 2020.

However, our analysis demonstrates that the notion of obscenity is not immanent within the imagery itself (e.g., revealing, or depicting certain body parts) nor in a victim's conduct, but rather homosocially and arbitrarily ascribed to gendered bodies as an oppressive control mechanism. We bring our focus to the way obscenity is fabricated and co-constructed through social interactions between perpetrators mediated by technological capacities feasible in each era. As homosocial bonding intersects with various technologies enabling social engagement, commenting, re-sharing, and image capturing/generation, perpetrators further augment \textit{vision} as power \cite{mulveyVisualPleasureNarrative1975} and collectively practice ascription of obscenity and voyeuristic male-gaze. 

In Era 1, obscenity was arbitrarily ascribed to the victim's personally identifiable traits (and any women who shared these traits) by spectators and sharers. Their traits (e.g., a red muffler or the victims' initials of their names) were utilized among male spectators across online and offline spaces as an alias to refer to a particular sexually exploitative material \cite{jeon2021n, sora_kim_changes_2018}. The naming of these materials with aliases effectively drew attention and curiosity to the victims. In anonymous chatrooms in PC communication, individuals asked around and shared information about where to obtain the tapes, naturalizing the voyeuristic desires of a male-dominant audience ~\cite{sora_kim_changes_2018}. Later, these aliases were also used as a meme humiliating women with that trait, as several videotapes got viral through word of mouth, and mainstream media uncritically adopted a few of these terms \cite{muffler1997, muffler1997sentence}. The harm of such obscenity labeling is that it covertly dissipated public attention towards who the victims were and obscured its abusive nature. Along with lurid reporting of mainstream media, these recordings of sexual violence were easily rendered as mere “sensual” pornography which featured “obscene” behaviors of promiscuous girls \cite{muffler1997}. The act of watching these videotapes together with close male peers formulated an early day cultural memory of male bonding.  

In Era 2, the framing of obscenity was collectively fabricated within online file-sharing repositories, such as online forum-based communities and Webhard storage. The lens of obscenity here is manifested and reified through the combined usage of (1) neologism and (2) spycam footage. The extensive use of alias characterized in the previous era transcended into neologism of hateful stereotypes of women or South Korean women in general, like the example of “Kimchinyeon”. Such textual expressions were usually accompanied with spycam footage uploads featuring random women in publicly accessible spaces. While neologism symbolically labeled imagined contextual obscenity, nonconsensual filmings materialized unsolicited projection of male-gaze onto women's bodies. Since the mainstream modality of visual medium transformed from videotapes in Era 1 to digital files in Era 2, online portal sites offered male-dominant “playgrounds” to consume and create DSV materials, alone but together ~\cite{sora_kim_changes_2018}. The forum-based platform layout and commentary section allowed them to collectively upload the abusive materials, label them, and leave comments on the featured bodies together. Higher-ranking creators were rewarded with high hit rates, comment counts and gained recognition among other members in Soranet as ``pioneers''~\cite{mtoday_soranet_2015}. In the meantime, the commentary section served as a breeding ground for members to practice objectifying women’s bodies as “obscene” materials, collaboratively interpreting and assessing their physical characteristics. Like this, the platforms functioned more than just a repository. It operated as a socializing space where the meaning of obscenity was socioculturally and collaboratively constructed, thus leading to justification of victimization \cite{kimjinsook2018misogyny, sora_kim_changes_2018}.


Enabled by encrypted messaging services and widespread use of social media services, we could observe cases wherein obscenity was coercively and literally attributed on victims. As opposed to the practice of light deanonymization leveraging identifiers of featured victims in Era 1, the male-gaze fantasy of mocking and harassing “obscene” women was executed through directly involving de-anonymized victims. Perpetrators in this era intentionally targeted vulnerable minors who run “deviant accounts” on social media and invited them by blackmailing and threatening of identity exposure to the encrypted Telegram chatrooms. These chatrooms were joined by a small number of victims and male participants, effectively functioned as a virtual “Panopticon” where they could collectively surveil and witness how the vulnerable victims were controlled by the threat of deanonymization and nonconsensual distribution. This difference in anonymity between the male-dominant audience and the female victims perpetuated the asymmetrical power relations underlying the male gaze. In this gaze, the sexualized fetish of spectators was imposed on the women who are bound to bear the gaze as still objects, while the voyeurs gain the power of looking, as well as a satisfying sense of omnipotence~\cite{mulveyVisualPleasureNarrative1975}. The victims are entrapped by the forcible commands of producing and sharing sexualized materials “voluntarily” and the social stigma of obscenity.

The advent of AI-based image generation technology in Era 4 set the scene for perpetrators to fully automate the materialization of the male gaze and the projection of obscenity onto women's bodies. Even when victims’ footage is not captured in explicitly sexual contexts-such as photographs from social media profiles-or when only personal information of female acquaintances (e.g., affiliation or home address) is available, these materials are readily overlaid with sexual imagery and transformed into obscene artifacts through AI generation. In this process, victims’ bodies and personal identities are decontextualized and fragmentized as modular inputs that are readily usable for the artificial fabrication of sexualized content. The automation of fabrication significantly lowers the threshold for creating abusive materials and participating in DSV practices, enabling the humiliation of female peer acquaintances as long as they have a screenshot of the target’s social media profile. As in Eras 2 and 3, the distinction between creators and consumers of abusive materials becomes increasingly blurred in this era. The humiliation of victims with deepfaked imagery further operates as an oppressive mechanism of control, particularly when coupled with threats of dissemination across social media platforms with victims’ real names attached. Again, through deanonymization, the subjugation of victims continues in effect as the enactment of the male gaze.

\subsection{Imperceptibility of Violence in Digital Sexual Violence}  \label{sec:findings-2}
Consent is a core concept in discussions of sexual violence, particularly within legislative and legal frameworks~\cite{westen2017logic, wertheimer2003consent, Lee2025SV, Choi2021nonconsent}. In Era 3 of our genealogy, victims targeted through “deviant accounts”, were treated as if they consented to DSV \textit{voluntarily}, in that victims had produced or shared images of themselves to the perpetrators. However, this presumes that the subject was situated in a moment where they could have noticed the harm and could have said `no' was realistically possible. Such an assumption overlooks the structural and situational conditions that constrain the victim's choice and agency. Moreover, it produces a backfire effect toward victims by retroactively placing the burden on them to explain why they did not expect or actively resist the harm. In this way, understanding consent and non-consent through voluntariness obscures relations of power and legitimizes the subsequent deprivation of victims’ agency. 

The case of deviant accounts illustrates these dynamics concretely: victims targeted through deviant accounts were disproportionately minors and marginalized women who are subjected to escalating threats, blackmail, and coercion~\cite{telegram_punished_2019, telegram_2019_woman, baksabaksa}---conditions that render any notion of ``voluntary'' participation fundamentally fictitious. Yet the voluntariness framing has produced concrete harmful consequences: legally, victims have been treated as accomplices in the distribution of obscene materials or child sexual exploitation content, facing criminal liability for their own victimization~\cite{muffler1997sentence}. 

Beyond the limitations of voluntariness-based interpretations, we argue that the consent framework itself is categorically inadequate for addressing DSV. Framing DSV as a matter of consent implicitly suggests that these acts could be legitimized if consent were present. Acts such as creating sexually explicit deepfakes or distributing intimate images constitute violations that are inherently harmful, regardless of whether the images were shared voluntarily. The question of consent, in this sense, is a distraction from the more fundamental issue: these are acts of violence that technology has rendered increasingly imperceptible to victims.

Our genealogical analysis across four eras reveals that technology has progressively transformed DSV into a form of violence that operates beyond the threshold of victims' awareness.  Rather than complicating consent, technological developments have created conditions under which violence occurs without perceptible moments---where harm is initiated, progresses, and intensifies in ways that victims cannot detect, anticipante, or respond to. We therefore shift our analytical focus from the concept of consent to the imperceptibility of violence, examining how evolving technologies systematically foreclose victims' ability to perceive that harm is occurring at all.


Victims are increasingly subjected to violence under conditions of multilayered imperceptibility, where harm unfolds in ways that are operationally undetectable. One key dimension of this imperceptibility concerns whether harm is even occurring, as victims may be unaware both of physical acts of filming and of the potential for non-sexual content to be repurposed into sexualized material. In Era 1, the violence was perceptible at its initiation: filming was carried out visibly, with victims aware that harm was being initiated, even though subsequent circulation remained unknown. This perceptibility at the initiation of harm became increasingly foreclosed with the introduction of portable and subminiature recording devices in Era 2. Filming became imperceptible, rendering physical coercion unnecessary and immediate awareness impossible while women’s bodies were digitized without their knowledge. In Era 4, deepfake and generative AI technologies expand conditions of imperceptibility by vastly broadening the range of materials that can be mobilized for the production of sexual imagery, such that any form of bodily representation, including everyday photographs or publicly available digital traces, can be repurposed into sexualized content. As a result, victims cannot perceive not only of when or where they are being recorded, but also which aspects of their digital presence can be transformed into sites of DSV.

Another key dimension of imperceptibility concerns how harm evolves and intensifies through the sequential and networked deployment of multiple technologies and practices, rendering the progression of violence invisible to victims. In Era 1, perpetrators directly filming victims and sharing images or videos made the violence relatively traceable. In Era 2, the number of perpetrators involved in DSV expanded through the intersection of multiple technologies, including file sharing platforms. Acts of filming, storing, and circulating images or videos became distributed across different technological infrastructures and actors, making the ongoing violence increasingly imperceptible to victims while adding new layers to their unknowing victimization. From Era 3 onward, perpetrators increasingly mobilized data from social media platforms, positioning these platforms as central actors in DSV and further obscuring the violence from victims' view. In Era 4, deepfake and generative AI technologies further intensify this condition by introducing fundamental opacity regarding how the victim’s images are processed, recombined, and transformed by AI systems. Such systems are trained on 
large scale public datasets whose subjects often lack knowledge of how their data may be reused, 
and victims have no way of knowing whether their images are retained, for how long, or across how many systems. 

This technological imperceptibility is compouneded by the temporal structure of DSV, which creates a gap between when violence occurs and when---if ever---it is perceived. Unlike physical violence experienced at the moment of its occurrence, technology-enabled DSV can persist and intensify over extended periods without the victim's knowledge---violence that exists not as a discrete event but as an ongoing condition. The longer victims remain unaware, the more extensively the harm propagates; in extreme cases, victims may never learn they have been victimized at all. This possibility of perpetual, unknowable victimization represents a technologically constituted form of violence. 

The imperceptibility of violence is further intensified through intersecting and evolving perpetrator practices that operate across the procedural stages of DSV, extending well beyond acts of acquisition and sharing. From Era 2, perpetrator involvement diversified and increasingly shifted from individual consumption to networked and communal participation, introducing collective forms of violence that are even less perceptible to victims. The impact of these networked structures is clearly illustrated by Soranet as a portal and community website, whose design facilitated collective engagement with nonconsensually uploaded content through norms of ``critique'' and bodily evaluation \cite{kim2018together, ha2017click}. Through such practices, violence became embedded not in a single perceptible act but in ongoing, participatory processes that normalized objectification and victimization while rendering harms diffuse and invisible. Similar patterns of collective sharing and commentary can be observed across the Nth Room, peer humiliation practices, and deepfake communities in Eras 3 and 4. The violence is thus doubly imperceptible: victims are unaware of specific acts of violation and unaware that entire communities have formed around their victimization.

The failure to adequately account for these conditions of imperceptibility produces significant limitations in policy responses to DSV. Many existing legal and regulatory frameworks continue to rely on narrow, moment based understandings of consent that presume victims’ awareness of harm, identifiable moments of non-consent, and clearly bounded acts of violation. South Korea's \textit{Act on Special Cases Concerning the Punishment of Sexual Crimes, Article 14}~\cite{korea_legislation_research_institute_klri_act_2016}, which criminalizes filming using cameras or similar devices ``against the will of the person being filmed,'' exemplifies this limitation. More broadly, the Korean legal system has consistently centered consent as the primary criterion for determining whether DSV has occurred and for assigning culpability. Courts and prosecutors have routinely examined whether victims were aware of filming, whether they expressed objection, and whether their ``will'' was violated at identifiable moments. This concent-centric approach has shaped not only adjudication but also public discourse, placing the burden on victims to demonstrate non-consent rather than treating the acts themselves as inherently criminal. Such formulations presuppose a moment at which the victim could have been aware of the filming and could have expressed their will---a presupposition that technology has systematically undermined. As the preceding analysis demonstrates, victims are often unable to recognize when harm begins, how it unfolds, or which actors and technologies are involved, particularly in cases involving covert recording, secondary circulation, platform mediated participation, and AI-based reprocessing. Policies that frame DSV through the lens of consent therefore misapprehend the nature of the violence they seek to address. By implicitly asking when and why consent was not withdrawn, such frameworks demand awareness from victims who have been structurally denied the possibility of awareness itself.

\subsection{Trading Abuse: Large-Scale Commercialization and Industrialization of Gender-based Violence} \label{sec:findings-3}

Across three decades, image-based DSV has evolved into a durable economic system in which digitized women’s bodies and acts of gender-based violence are rendered tradable. 
DSV has evolved into what could undoubtedly be called an industry. The analyzed cases denote that perpetrators have constructed systematic and decentralized revenue structures around transactions of nonconsensual sexualized images with appropriation of various technological apparatuses such as digital currency, encrypted messaging, and generative AI. 

Initially, image-based DSV did generate monetary profits for a few individual perpetrators, yet it was difficult to call it an industry. As seen in the early incident in Era 1, where in underage perpetrators sold nonconsensually produced videotapes to their immediate peers, small-scale firsthand monetary exchanges took place. Direct sales were made between producers and consumers through online or offline communication, followed by physical handover. While these practices were not yet industrialized, early signs of commercialization were already present when pirated copies of such materials entered private vendors of electronics shopping centers like Sewoon marketplace \cite{muffler1997}. Visual materials of sexual extortion began to be traded by retail sellers beyond original creators themselves, indicating emerging demand for sexualized content framed by its “realness” and proximity to everyday life.

In Era 2, firsthand interindividual transactions of nonconsensual images still persisted as can be seen in 2015 Waterpark Molka Crimes. However, in this era, the monetary ecosystem of image-based DSV arguably went through a substantial expansion. The Webhard Cartel case exemplifies how profit making emerged as one of the prominent motivations of committing, co-conspiring, and cooperating with DSV. Webhards and widespread distribution of mobile phones, as sociotechnical infrastructures, enabled the formulation of a transactional retail network of nonconsensually taken image/videos \cite{webhardcartel_2025}. From this era onward, actors involved in DSV diversified, such as direct actors (producers/filmers, uploaders, individual consumers, and platform owners) and indirect beneficiaries (owners of illegal business such as sex trafficking or gambling and digital undertakers). Platform owners played a key role in collecting and distributing profit, drawn from individual consumers' purchase of in-platform virtual currency (e.g., points), to producers/uploaders/consumers. Notably, monetary exchange was rather centralized through the use of in-platform currency. Users needed virtual money to download image/videos, clearly marking individuals' transactional history. Platform owners supplied this money as incentives to members who sustained in-platform engagement (e.g, friend invitation) or made continuous uploads, which in turn led to more downloads and widespread distribution of these contents. Our analysis contends that this establishment of transactional DSV network set out the scene for a large-scale commercialization of nonconsensual sexual images in South Korea and persists at a broader level until nowadays, even after the official shutdown of Soranet and Webhards.  

As the image-based DSV industry entered the third era, its transactional network became more decentralized, intertwined with diverse profit-driven actors and technological scaffolding. Unlike the previous era, nonconsensual image/videos were traded and distributed in multiple interconnected anonymous Telegram chatrooms as illustrated in Nth Room Case. Access to these chatrooms were rather openly advertized in mainstream social media platforms, expanding their consumer range into a national-scale. The operation of these chatrooms employed organized business schemes such as freemium plans, auctions, and crowdsourcing, through which other types of criminally abusive acts were palpably encouraged and committed in a participatory manner. In the following, we illustrate how decentralized transactional mechanisms fueled rhizomatic system of harm \cite{johnsonDeepfakeViolenceFuture}, in which gender-based harm spreads through digital systems non-linearly, without clear beginning nor end. 

Pseudonymous administrators of interconnected parallel Telegram chatrooms in Era 3 and 4 collectively implemented Freemium tactics. They first offered free entry and sample image/videos to aggregate as many free members as possible and then attracted them to pay a premium for more exclusive contents. This practice is evident in the way Nth Room administrators induced free members' conversion into paid users by occasionally posting free sexual abuse content in entry chatrooms \cite{park_hani_2022}. Similar tactic was used in the Deepfake Case where chatroom participants who wanted to use genAI bots more than freebies purchased and made payment with cryptocurrency to chatroom admins. Such approach made a lucrative business. For instance, administrators of a paid Nth Room, with 260,000 active users, reportedly had made several billion KRW worth revenues (approx. several million USD). They made profit by differentiating the entry pricing of various chatrooms, charging more money for those rooms with allegedly \textit{more exclusive and rare} contents corresponding to the severity of abuse, humiliation, and, revealingness of imagery. Some of these rooms charged up to 1.5 million KRW (approx. 1000 USD) for entry \cite{park_hani_2022}. Furthermore, chatroom administrators leveraged many other tactics, such as crowdsourcing or auctions \cite{military2024, park_hani_2022}. What is even more appalling is that their revenue schemes had worked out. This implies a horrid reality where DSV and misogynic humiliation had become a consumable sociocultural content persistently in demand \cite{park2024deepfake22}. 

In-group members' voluntary conspiracy in criminal and abusive acts further strengthened these rhizomatic monetary transactions of DSV, formulating a loyal guild of accomplices. Chatroom members' execution of criminal or abusive acts were encouraged and incentivized as rite of passage or valid reward that is worth equivalent to real money. As described in Era 3, in these rooms, nonconsensual footage of women's bodies and screenshots or photos verifying that one had perpetrated online/offline sexual harassment or abuse were often received as a monetary alternative of an entry/download/generation fee \cite{jo_hani_21, nthroom_baksa, donga_deepfake2024, military2024}. To access rarer and more exclusive materials, members engaged in generating and circulating the content, willingly becoming complicit with the system of abuse. Vice versa was observable as well, members paying to be complicit. For instance, in the Nth Room case, some members purchased a special ‘voucher’ from the room admin that gives them a priority to command demeaning actions to victims forcibly invited to a livestreamed sextortion chatroom \cite{nthroom_baksa}. Within this malicious ecosystem where commitment of sexual abuse and monetary reward were deemed interchangeable, the boundary between producers and consumers became gradually blurry. The use of cryptocurrency and end-to-end encrypted message services further reinforce these transactions by making them harder to be trackable. In this way, newcomers soon turn into loyal consumers, promoters, co-producers, and in the end fervent producers themselves, completely clouding the sense of guilt conscience. 

On top of this, the adoption of generative AI technologies, like nudifying applications or deepfake bots, considerably contributed to the industry's expansion by streamlining the harm. Generation of nonconsensually sexualized image/videos became not only easier, faster, and accessible to laypeople, but also cheaper and frictionless. These technologies in Era 4 automated the coercive practices of sextortion carried out by humans in Era 3. In Nth Room case, perpetrators carried out blackmailing and sextortion to capture and obtain free image/videos of victims, which they re-shared in other rooms for profit \cite{telegram_2019_woman}. Or else, a few skilled members manually used Photoshop to create sexualized images and prompted other members to make payments. While these types of abuse continue to occur in Era 3, with the use of genAI, perpetrators can now generate custom-sexualized image/videos without risking exposing themselves, sparing labor and time to harvest and produce imagery. Anyone could generate a nudified image of a person in 5 seconds with only 650 KRW (0.49 USD) \cite{park2024deepfake22}. A mass production of sexually demeaning image/videos became feasible, exemplified by a case of an individual perpetrator who generated 600 deepfake pornography synthesizing k-pop idols' faces with kid models' bodies \cite{newsis_2025}. In comparison to Era 2, where the scale of harm rather proportionally corresponded to the volume of sexual content uploaded and circulated by human contributors, with generative AI systems, a single, non-sexual, publicly available personal image can be transformed into sexualized contents with minimal effort, and that content could then be repeatedly reproduced, modified, reused, and redistributed at scale~\cite{ntoday_114563}. Furthermore, these deepfake bots further anonymize perpetrators as they automate the collection/retrieval of download fees \cite{donga_deepfake2024}, rendering human-human conversations on monetary transactions redundant. This makes it more difficult for authorities to identify and investigate transactional trajectories without cooperation from Telegram \cite{park2024deepfake, fnnews_202409101825398415}. 

\section{Discussion}

\subsection{Implications for CSCW and HCI Research on DSV}

Here, we discuss implications of our findings for future CSCW and HCI research. First, we show how obscenity is fabricated through male homo-social communities, wherein victims' imagery becomes collectively framed as obscene through participatory practices in male-dominant networks. In DSV contexts, obscenity is arbitrarily ascribed to victims through collective engagement of imagery to win peer recognition, visibility, and exert power, rather than through the visual properties of the images themselves. This perspective suggests that future work should reorient its focus away from defining what obscene, sexual, or intimate imagery is towards how such connotation is performatively and relationally constructed. Our findings reveal that homosocial communities function as a central engine in this process. Accordingly, we propose that research on DSV can benefit from scholarship on male-dominant communities which examines how they are maintained, internal norms are produced and reinforced, and gendered power relations are reproduced through collective practices~\cite{massanari2017gamergate, rubin2020fragile}. By foregrounding homosocial fabrication as a core cultural mechanism, future work can move beyond content-level explanations to examine gendered sociocultural structures that sustain DSV more deeply, opening up a space for more culture-sensitive and structurally informed interventions.

Second, we demonstrate how DSV has become increasingly imperceptible as networks of multiple technologies intersect with evolving and diversified perpetrator practices, thereby eroding the conditions under which harm can be recognized, anticipated, or contested. Victims are often unable to perceive when harm occurs, predict its future trajectories, or meaningfully intervene. This imperceptibility poses a fundamental challenge to consent-based approaches that have long structured HCI and CSCW resarch on DSV, particularly image-based sexual abuse~\cite{zytko2022consent,qiwei_feminist_2024,qiwei2024sociotechnical,strengers2021can,qin2024did}. While previous studies have made important contributions by designing systems that facilitate consent or enable individuals to manage consent over their digital content, our genealogical analysis demonstrates that such approaches become inadequate when violence itself unfolds beyond victims' awareness. In these contexts, consent cannot be meaningfully granted, withheld, or revoked. As Barocas and Nissenbaum argue~\cite{barocas2009notice}, notice-and-consent frameworks often function as symbolic gestures that obscure deeper structural problems, shifting responsibility onto individuals while leaving the very enabling infrastructures intact. We therefore argue that the increasing imperceptibility of DSV necessitates a rethinking of consent as an analytical and design framework, calling on HCI and CSCW research to move beyond consent as a checkbox toward reimagining meaningful forms of protection in evolving technological systems.

Third, we explicate how image-based DSV has become an organized industry enabled by decentralized and rhizomatic monetary and criminal networks. Barriers to participation diminish and the distinction between operators, accomplices, and consumers gets blurred as DSV becomes perceived as purchasable product. Such dynamic allows DSV to propagate non-linearly, without a clear origin or endpoint: even when specific nodes are dismantled-such as the shutdown of the N-th Room-similar practices re-emerge in intensified or replicated forms, as observed in the subsequent Doctor’s Room~\cite{park_hani_2022} and later deepfake cases where then participants transitioned into operators~\cite{donga_deepfake2024}. Because rhizomatic harm persists even when visible branches are removed, punitive approaches such as content takedowns, platform bans, and criminal prosecution, while necessary, are insufficient on their own. In line with prior scholarship that characterizes nonconsensual sexual imagery as rhizomatic and thus requiring holistic, relational responses~\cite{dodge2022restorative}, CSCW and HCI research could consider restorative approaches that prioritize accountability, repair, and community-based recovery over purely punitive measures~\cite{schoenebeck2021youth,10.1145/3555775,10.1145/3544548.3581512}.

\subsection{Reflections on Genealogical Approach}

In this paper, we adopted a sociotechnical genealogical approach to examine DSV in South Korea. Here, we reflect on what this methodological choice enabled and what it offers for future CSCW and HCI research on digital harms. 

Genealogy allowed us to move beyond treating DSV as a series of isolated incidents or as a problem reducible to technological misuse. Dominant framings of DSV---whether in policy, media, or research---often focus on specific technologies (spycams, deepfakes) or specific cases (Nth room, Telegram scandals) as discrete problems requiring targeted solutions. A genealogical approach instead enabled us to position these incidents within longer historical trajectories, revealing how they are shaped by and, in turn, reproduced enduring sociocultural formations~\cite{jang2025expanding}. By tracing DSV across four eras and three decades, we could identify not only what changed but also what persisted: the underlying misogyny, the homo-social dynamics of perpetration, and the structural inadequacies of legal and institutional responses.

Central to our genealogical analysis is attention to both \textit{continuity} and \textit{rupture}~\cite{jacobsen_tensions_2024}. Rather than viewing technological changes as a linear sequence of innovations, genealogy foregrounds how sociotechnical arrangements persist, transform, and re-emerge over time. As Jacobsen and Simpson argue, emerging technologies like deepfakes ``should be seen as embedded in a much larger assemblage of events and powers that were in place long before'' they emerged~\cite[p.1098]{jacobsen_tensions_2024}. This framing proved essential for our analysis. The continuities we identified---such as the collective construction of ``obscenity'' through male-dominant networks---reveal how longstanding patriarchal practices persist through new technological forms. At the same time, the ruptures we identified---the increasing imperceptibility of violence, the industrialization of abuse---highlight how technology reconfigures the scale, speed, and character of harm. Crucially, genealogy helped us see that these ruptures do not replace but rather layer upon earlier formations: Era 4's automated deepfake production builds on Era 2's transactional logics; AI-enabled imperceptibility extends the covert surveillance normalized through spycam culture.


This genealogical approach has several implications for CSCW and HCI research. First, it challenges the tendency toward reactive, technology-specific interventions. Dominant responses to DSV---detection algorithms, content moderation, platform policies---often treat each new technological threat as a novel problem requiring novel solutions. Our analysis suggests that such approaches, while necessary, are insufficient because they leave unexamined the sociocultural continuities that sustain harm across technological change. As Jacobsen and Simpson observe, ``these issues cannot be solved through deepfake detection techniques and algorithms alone''~\cite[p.1106]{jacobsen_tensions_2024}. Second, genealogy offers a method for understanding why certain configurations of harm persist despite repeated interventions. By attending to historical formations, researchers can identify the conditions that enable harm to be reconstituted in new forms---and thus design interventions that address root causes rather than symptoms. Third, genealogy foregrounds the situatedness of digital harms. Our analysis is grounded in the specific sociocultural context of South Korea, yet the approach itself---tracing how local histories, cultural norms, and technological affordances co-constitute harm---can be adapted to other contexts. We encourage future research to develop genealogies of DSV and other digital harms in different cultural settings, attending to both local specificities and cross-cultural patterns. Therefore, genealogy functions as both an analytic and generative method: it exposes the historical contingencies underlying present conditions while opening space for interventions that address the continuities of power even as technologies transform.

\section{Conclusion}
In October 2024, following the South Korea deepfake scandals, we convened as a collective of HCI researchers compelled to act. Despite our varied research interests within HCI, a shared sense of urgency brought us together. 
An activist from the organization ReSET, which has tracked and documented Telegram-facilitated sexual exploitation since 2019, remarked~\cite{nam_2024}:
\textit{``While things may seem quiet for now, they are likely still active on overseas platforms because digital spaces have no borders. People tend to view digital sexual crimes as issues that exist only when they’re in the headlines, but these crimes occur every single day.''} 
This paper calls on HCI and CSCW researchers to recognize and engage with the persistent and “everyday” nature of technology-facilitated “suffering.” While this describes the history of DSV from the 1990s to present day, the story is very much still ongoing. 
On January 2025, another harrowing case was reported: a Telegram group with over 200 members producing deepfake sexual exploitation material of teenagers~\cite{jan23_2025}. 
South Korea holds the unenviable title of the country most targeted by deepfake pornography, and its impact permeates the day-to-day and personal lives of women. As such, we emphasize the importance of our critical analysis situated in the specific context of South Korea, with the hope that our analysis can inform responses and prevention efforts for similar crimes in the future. We hope that other HCI researchers join us in reflecting on what our next steps could be and, empathetically and critically, explore ways to intervene and address this ongoing crisis, both in Korea and internationally.

\begin{acks}
We express sincere gratitude to feminist activists and investigative journalists in South Korea, such as Team Flame, for their brave work, without which this work would not have been possible. Thanks to Rob Comber at MID, KTH and Creating Ethics Infrastructure lab members at Georgia Tech for feedback and support. We are grateful to anonymous reviewers for their valuable feedback.

This work was co-funded by the European Union (ERC, Intimate Touch, 101043637). Views and opinions expressed are however those of the author(s) only and do not necessarily reflect those of the European Union nor the European Research Council and other funding bodies. Neither the European Union nor the granting authority can be held responsible for them. 
\end{acks}

\bibliographystyle{ACM-Reference-Format}
\bibliography{dkorea,reference}

\end{CJK}
\end{document}